\documentclass{llncs}

\usepackage[utf8]{inputenc}
\usepackage{color}
\usepackage{listings}
\usepackage{graphicx}
\usepackage{array}
\usepackage{textcomp}
\usepackage{xspace}
\usepackage{booktabs}
\usepackage{pifont}
\usepackage{hyperref}
\usepackage[numbers,sort,square]{natbib}
\usepackage{enumitem}
\newcounter{descriptcount}
\usepackage[nolist]{acronym}

\renewcommand*{\acsfont}[1]{\textbf{#1}}
\renewcommand*{\acffont}[1]{\textbf{#1}}
\renewcommand*{\acfsfont}[1]{\textbf{#1}}

\makeatletter
\renewcommand*{\@acf}[1]{%
  \ifAC@footnote
    \acsfont{\AC@acs{#1}}%
    \footnote{\AC@placelabel{#1}\hskip\z@\AC@acl{#1}{}}%
  \else
    \acffont{%
      \acfsfont{\acsfont{\AC@acs{#1}}}%
      \nolinebreak[3] %
      (\AC@placelabel{#1}\hskip\z@\AC@acl{#1})%
    }%
  \fi
  \ifAC@starred\else\AC@logged{#1}\fi}
\makeatother

\newlist{enumdescript}{description}{1}
\setlist[enumdescript,1]{%
  before={\setcounter{descriptcount}{0}%
          \renewcommand*\thedescriptcount{\arabic{descriptcount}}}
  ,font=\bfseries\stepcounter{descriptcount}\thedescriptcount.~
}

\newcommand{\metric}{metric\xspace}
\newcommand{\metrics}{metrics\xspace}
\newcommand{\thesystemtitle}{XSS Peeker}
\newcommand{\thesystem}{\textsc{\thesystemtitle}\xspace}

\newcommand*\OK{\ding{51}}
\newcommand*\KO{\ding{55}}

\usepackage{soul}
\usepackage[disable]{todonotes}

\usepackage{microtype}
\usepackage{enumitem}
\usepackage[labelfont=bf,belowskip=1pt,aboveskip=4pt,skip=1pt,margin=0pt,font=small]{caption}
\let\llncssubparagraph\subparagraph
\let\subparagraph\paragraph
\usepackage[small,noindentafter]{titlesec}
\let\subparagraph\llncssubparagraph
\titlespacing{\section}{0pt}{*1.8}{*1.2}
\titlespacing{\subsection}{0pt}{*1.5}{*0.8}
\titlespacing{\subsubsection}{0pt}{*1.2}{*0.4}
\titlespacing{\paragraph}{0pt}{*0.6}{*0.5}
\setlist{leftmargin=12pt,itemsep=0.2pt,topsep=2pt}
\setitemize{leftmargin=12pt,itemsep=0.2pt,topsep=2pt}
\setenumerate{leftmargin=12pt,itemsep=0.2pt,topsep=2pt}
\begin{document}

\begin{acronym}
  \acro{phase1}[Phase~1]{Payload Extraction}
  \acro{phase2}[Phase~2]{Payload Templating}
  \acro{phase3}[Phase~3]{Template Evaluation}
  \acro{phase4}[Phase~4]{Retrofitting Negative Payloads}

  \acro{length}[M1]{Length}
  \acro{charset}[M2]{Number of distinct characters}
  \acro{callbacks}[M3]{Custom callbacks}
  \acro{encodings}[M4]{Multiple encodings}
  \acro{evasion}[M5]{Number of known filter-evasion techniques}
  \acro{lev}[M6]{Mean and variance of Levenshtein distance}
\end{acronym}

\title{\thesystemtitle: A Systematic Analysis of Cross-site Scripting
  Vulnerability Scanners}

\titlerunning{\thesystemtitle}

\author{Enrico Bazzoli\inst{1} \and
  Claudio Criscione\inst{1,}\inst{2} \and
  Federico Maggi\inst{1} \and
  Stefano Zanero\inst{1}}

\institute{DEIB -- Politecnico di Milano, Italy,
  \email{enrico.bazzoli@mail.polimi.it},\email{\{federico.maggi,stefano.zanero\}
    @polimi.it} \and Google, Zurich, Switzerland, \email{claudioc@google.com} }

\authorrunning{Enrico Bazzoli et al.}

\tocauthor{Enrico Bazzoli (Politecnico di Milano),
  Claudio Criscione (Google),
  Federico Maggi (Politecnico di Milano),
  Stefano Zanero (Politecnico di Milano)}

\maketitle\vspace*{-0.5cm}

\begin{abstract}
  Since the first publication of the ``OWASP Top 10'' (2004), cross-site
  scripting (XSS) vulnerabilities have always been among the top 5 web
  application security bugs. Black-box vulnerability scanners are widely used
  in the industry to reproduce (XSS) attacks automatically. In spite of the
  technical sophistication and advancement, previous work showed that black-box
  scanners miss a non-negligible portion of vulnerabilities, and report
  non-existing, non-exploitable or uninteresting vulnerabilities. Unfortunately,
  these results hold true even for XSS vulnerabilities, which are relatively
  simple to trigger if compared, for instance, to  logic flaws.

  ~~~Black-box scanners have not been studied in depth on this vertical: knowing
  precisely how scanners try to detect XSS can provide useful insights to
  understand their limitations, to design better detection
  methods. In this paper, we present and discuss the results of a detailed and
  systematic study on 6 black-box web scanners (both proprietary and open
  source) that we conducted in coordination with the respective vendors. To
  this end, we developed an automated tool to (1) extract the payloads used by
  each scanner, (2) distill the ``templates'' that have originated each payload,
  (3) evaluate them according to quality indicators, and (4) perform a
  cross-scanner analysis. Unlike previous work, our testbed
  application, which contains a large set of XSS vulnerabilities,
  including DOM XSS, was gradually retrofitted to accomodate for the payloads
  that triggered no vulnerabilities.

  ~~~Our analysis reveals a highly fragmented scenario. Scanners exhibit a wide
  variety of distinct payloads, a non-uniform approach to fuzzing and mutating
  the payloads, and a very diverse detection effectiveness. Moreover, we found
  remarkable discrepancies in the type and structure of payloads, from complex
  attack strings that tackle rare corner cases, to basic payloads able to
  trigger only the simplest vulnerabilities. Although some scanners exhibited
  context awareness to some extent, the majority do not optimize the choice of
  payloads.
\end{abstract}

\section{Introduction}
\label{sec:introduction}
The software industry is increasingly paying attention to security, with high
profile incidents being more and more frequently leading news in the media.
Nevertheless, web applications often contain flaws that,
when successfully exploited, lead to dangerous security breaches. Web
application vulnerabilities are one of the most important and popular security
issues, constantly making it to the top
list of disclosed vulnerabilities~\cite{xforce,dell}.

Cross-site scripting (also known as XSS) is a prevalent class of web
application vulnerabilities~\cite{xssTop10}, despite being well known and
studied in depth. Reported cases of XSS have increased significantly
between 2004 and 2013~\cite{msrcXss,Cenzic}. In June 2013, XSS was
reported as the most prevalent class of input validation vulnerabilities
by far~\cite{Tudor2013}.

Black-box scanners are easy to use tools used to check web applications
for vulnerabilities, including XSSs. It has been shown that black-box scanners
miss a non-negligible
portion of vulnerabilities, i.e. that they display false
negatives~\cite{Doupe2010,Suto2010a,Vieira2009}, and that symmetrically they
often report non-existing, non-exploitable or uninteresting vulnerabilities,
i.e. they exhibit false positives~\cite{Vieira2009}. Although XSS
vulnerabilities are relatively easy to discover (for instance if compared with
flaws in business logic), previous work and our own pilot study showed that
black-box scanners exhibit shortcomings even in the case of XSS flaws.

To our knowledge (see \S\ref{sec:related-work}) there is no detailed study
of black-box web vulnerability scanners that focused specifically on XSS
vulnerabilities and their detection approaches.
Previous works and commercial benchmarks considered XSS bugs as part of a set of
flaws in testbed web applications, but not as the main focus. Also, it
is important to note that previous works measured the detection rate and
precision of the scanners mainly with the goal of benchmarking their relative
performance. Although we believe that these indicators are important, providing
precise
insights on the structure, generality, fuzzing mechanism, and overall quality of the XSS
detection approaches could help web application developers to design better
escaping and validation routines, and scanner vendors to \emph{understand}
the reasons behind scanner weaknesses. These goals require a different
perspective: rather than using a testbed web application with difficult-to-find
entry points, complex session mechanisms, etc., normally leveraged by testers
to challenge a scanner or by vendors to demonstrate their
product's sophistication, we
believe that the test cases should be easy to find and comprehensive, so to
maximize the
coverage of the analysis. Since the main detection approach for black box
testing for XSS is to fuzz HTTP requests by injecting ``payloads''
\cite{owaspTestingGuide},
and empirical evidence shows that scanners do the same, maximizing coverage
means, for our purposes, to observe the highest number of payloads.

We followed a simple methodology of our own devising, with the aid of an
automated tool that we
developed, \thesystem, which sniffs network packets, decoding the HTTP
layer and extracting the XSS payloads. In addition, \thesystem streamlines some
tedious tasks such as finding groups of related payloads, making
automatic analysis and result visualization feasible even in the case of large
amounts of payloads.

Since our objective is to obtain as many payloads as possible from the
scanners, we do not hide the vulnerable entry points in any way. In other
words, our results show a best-case scenario for each scanner.
Moreover, we designed our own testbed web application, which
we called \emph{Firing Range}\footnote{available online at
  \url{http://public-firing-range.appspot.com}}, in such a way that it is easy
to insert new
vulnerabilities. Whenever we detected a payload that was not triggering any of
our test cases, we reverse engineered a test case that would
satisfy specifically that payload. We applied the process iteratively,
running new scans and collecting new payloads. We also added specific test cases
for DOM XSS vulnerabilities, one of the most recently discovered and emerging
sub-classes of XSS flaws~\cite{Lekies2002,Heiderich2013}, not included in
previous works. It is worth noting that our research is not limited to
individual scanners: we
observed how cross-scanner analysis yields very interesting results.

\medskip\noindent In summary, in this paper we make the following contributions:
\begin{itemize}
\item A publicly available testbed web application that exposes non trivial XSS
  vulnerabilities (detailed in \S\ref{sec:firing-range}), augmented with
  the new cases discovered in our results.
\item A methodology to analyze how black-box web application scanners work by
  (1) extracting the payloads from the HTTP requests, (2) clustering them to
  scale down the evaluation challenge and keep it feasible, and (3) evaluating
  each cluster in terms of use of evasion, compactness, and other
  quality indicators.
\item A publicly available prototype implementation of the above
  methodology~\footnote{available online at
    \url{http://code.necst.it/xss-peeker}}
\item A detailed measurement applied to 6 scanners: Acunetix 8.0, NetSparker
  3.0.15.0, N-Stalker 10.13.11.28, NTOSpider 6.0.729, Skipfish 2.10b and w3af
  1.2.
\end{itemize}
For anonymity reasons, and because by no means we want to provide a benchmarking
analysis, the names of the scanners appear, in random order, as
\texttt{Scanner1}, \texttt{Scanner2}, etc., in our results.

\section{Background}
\label{sec:background}
Before describing our approach in detail, we introduce the essential background
concepts and terminology on web application XSS vulnerabilities and black-box
scanners.

\subsection{Cross-site Scripting Vulnerabilities and Attacks}
XSS \emph{attacks} consist in the execution of attacker-controlled code
(e.g., JavaScript) in the context of a vulnerable web application.
In this paper, we refer to the portion of malicious code as \emph{payload}.
XSS attacks are made possible by input-validation
flaws (i.e., XSS \emph{vulnerabilities}) in web applications, which accept
input from a user (i.e., attacker) and deliver it to the client (i.e., victim)
as part of an HTTP response. An attacker is thus able to deliver malicious code
to a victim, with consequences ranging from information or credential stealing
to full violation of the victim machine or network (e.g., by leveraging further
browser vulnerabilities). The attacker is able to access cookies, session
tokens and other sensitive informations, and also to rewrite the content of the
HTML page.

Without aiming for a complete taxonomy, XSS vulnerabilities and attacks
can be divided in \emph{stored} and \emph{reflected}, and both can be
\emph{DOM based} or not. In reflected attacks the victim is tricked
(e.g., through links or short URLs~\cite{longshore} embedded in
e-mail or instant messages) into sending a specially crafted request---which
embeds the actual payload, which is bounced back to the client immediately.
In stored XSS attacks the moment the payload is injected is decoupled
from the moment that it is effectively displayed and executed by the victim,
as the attacker's code achieves some form of persistence.
DOM-based attacks \cite{klein2005dom} can be distinguished as they rely
on the insecure handling of untrusted data through Javascript rather
than the simple reflection of a full payload.

\subsection{Black-Box Web Vulnerability Scanners}
Black-box web vulnerability scanners leverage a database of known security
attacks, including XSS payloads, and try to trigger and detect potential
vulnerabilities. More precisely, black-box scanners \emph{crawl} the target web
application to enumerate all reachable pages with \emph{entry points} (e.g.,
input fields, cookies), generate (mutations of) input strings based on their
database, inject the resulting payload in the entry points and analyze
application's responses using an \emph{oracle} to detect the presence of
vulnerabilities (e.g., by looking for the injected payload in the output).

Black-box scanners are, by definition, agnostic with respect to the application
internals and functioning, and thus virtually suitable for any application
based on common web stacks.

When evaluating or analyzing a black-box vulnerability scanner, a so-called
\emph{testbed web application} is used. A common claim made by vendors is that
testbed applications they provide closely mimic the relevant properties of
real world applications. As we will detail in the next section, this is not
our approach.

\section{Development of Our Testbed Web Application}
\label{sec:firing-range}
The implementation of the testbed web application is a key point. Given our
goals and needs, the requirements of such a testbed are: to have clearly defined
vulnerabilities, to be easily customizable, to contain the main types of XSS
vulnerabilities. Realistic, full-fledged testbed web applications have
been implemented in previous works~\cite{Doupe2010,Bau2010a,webGoat,hacmeBank},
but since they did not entirely meet our requirements, we chose to
implement our own application, which we called \emph{Firing Range}.

Meeting the above requirements, however, is tricky. On one hand, as pointed
out in~\cite{Doupe2010}, the complexity of some applications can hinder the
coverage of the scanner (e.g., if a vulnerable functionality is hidden behind a
complex JavaScript handler, the crawling component of the scanner might fail to
identify the vulnerable URL). On the other hand, there is no guarantee of
comprehensiveness of the testbed: vast portions of the assessed tool's
capabilities might remain unexposed as they activate only under specific
conditions that happen not to be triggered by the test application.

We decided to address these shortcomings explicitly while designing \emph{Firing Range}.
Our testbed offers no extraneous features, all links to the
vulnerabilities are explicitly exposed through HTML anchors and vulnerable
parameters are provided with sample input to improve discoverability.
This approach allows testing to be performed by providing the full list of
vulnerable URLs, thus removing any crawling-related failure entirely.
Furthermore, each vulnerable page is served as a standalone component with no
external resources such as images or scripts, as to create a minimal test case.

The main implementation challenge when creating a testbed is of course deciding
what tests it should include. We wanted our initial set to cover as many
cases as possible, but there was no such list readily available in previous
literature.
Since we basically wanted to test the detection of XSS on an HTML page,
and considering the HTML parsers in modern browsers have a finite set of states,
we resolved to create tests for each one of those.
To this end we analyzed the different contexts identified by the contextual
auto-escaper described in~\cite{Samuel}: simply reversing the perspective of a
parser, tasked with applying proper escaping to malicious payloads, provided us
with clear samples of the different contexts.
We generated test cases covering each of them with the
goal of (1) producing a ``vulnerable baseline'' of the different states of an
HTML parser and (2) inducing the scanners to inject as many payloads as
possible. Given the fairly limited amount of cases, we manually
generated tests where the input payload was echoed in the proper HTML context:
during this first seeding we did not include any kind of escaping or
filtering, and inputs were directly echoed back in the output page. It is worth
noting that there is no functional difference between
stored and reflected XSS when considering this perspective: simply changing the
origin of the input to be echoed from the URL to a persistent data source
allows to move from one type to the other. From the point of view of the
payload analysis, which is our core focus, they are thus indistinguishable from
one another. Therefore, for ease of development and of experimental
repeatability, our testbed web application only contains reflected
vulnerabilities.

HTML contexts are however not enough to generate test cases for DOM XSSs, which
exploits interactions between the DOM generated by parsing the original HTML
and JavaScript code. For DOM XSS, we started from the XSS
Wiki\footnote{\url{https://code.google.com/p/domxsswiki/wiki/Introduction}}, and other openly available collections of sample
vulnerabilities, and generated  a list of valid DOM sinks and
sources---which, notably, include sources other than URLs such as cookies
and browser storage. Each one of our DOM tests couples one of these sinks and sources.
All the \emph{Firing Range} tests have been manually verified as exploitable
with an appropriate payload or attack technique.

\subsection{Test Cases Examples}

Although the publicly available website of the testbed already provides full
details of the vulnerabilities we have tested, in this section we provide some
\emph{representative} examples for the sake of clarity.

\begin{lstlisting}[language=html,caption={DOM XSS from location.hash to innerHTML}.]
<html>
  <body>
    <script>
      var payload = window.location.hash.substr(1);
      var div = document.createElement('div');
      div.id = 'divEl';
      document.documentElement.appendChild(div);

      var divEl = document.getElementById('divEl');
      divEl.innerHTML = payload;
    </script>
  </body>
</html>
\end{lstlisting}

\begin{lstlisting}[language=html,caption={DOM XSS from \texttt{location} to \texttt{setTimeout}}]
<html>
  <body>
    <script>
      var payload = window.location;
      setTimeout('var a=a;' + payload, 1);
    </script>
  </body>
</html>
\end{lstlisting}


\begin{lstlisting}[language=html,caption={DOM XSS from \texttt{documentURI} to \texttt{document.write()}.}]
<html>
  <head><title>Address based DOM XSS</title></head>
  <body>
    <script>
      var payload = document.documentURI;
      document.write(payload);
    </script>
  </body>
</html>
\end{lstlisting}

\begin{lstlisting}[language=html,caption={Reflected XSS from \texttt{?q=payload}
  to a JavaScript slash-quoted assignment}.]
<html>
  <body>
    <script>var foo=/payload/;</script>
  </body>
</html>
\end{lstlisting}

\begin{lstlisting}[language=html,caption={DOM XSS from \texttt{window.name} to \texttt{eval()}.}]
<html>
  <head><title>Toxic DOM</title></head>
  <body>
    <script>
      var payload = window.name;
      eval(payload);
    </script>
  </body>
</html>
\end{lstlisting}

\begin{lstlisting}[language=html,caption={Reflected XSS from
\texttt{?q=payload} to \texttt{eval()}.}]
<html>
  <body>
    <a href="payload">Link!</a>
  </body>
</html>
\end{lstlisting}

\subsection{Iteratively discovered test cases}
During our experimental evaluation described in \S\ref{sec:experiments},
\thesystem discovered payloads that were not exercising any test case (i.e.,
vulnerability). Instead of limiting our analysis to report this discrepancy, we
iteratively constructed new test cases and progressively re-ran all the
scanners as described in \S~\ref{sec:negative-payloads}.

Reverse engineering the missing test cases was mostly a manual work thanks to
their limited number. Two of the payloads that did not exercise
any of our test cases in one of the early runs are the following:
\begin{lstlisting}[language=html,caption={Sample negative payloads.}]
  <div style="width:expression(alert('XSS'));">
  <script\%0d\%0a\%0d\%0a>alert(/xlqjgg4y/)</script>
\end{lstlisting}
Our goal was to design a reasonable test that would be ``discovered'' by these
payloads and not by others, while not overfitting to the point of
string-matching the payload. Our newly designed test case would allow the
injection of HTML tags as part of a payload, but would block any \verb|script|
tag and only allow \verb|style| attributes. Executing JavaScript in this case is
non trivial: a valid venue to exploit and thus detect the XSS vulnerability is
the one leveraged by the payload that suggested the test case, where the
\emph{expression} CSS property can be used to execute javascript in certain
browsers.

This iterative process produced 42 new test cases that were not identified by
our initial seeding. Consequently, this approach greatly improved the
testbed. For the sake of brevity, we refer the reader to
\url{http://public-firing-range.appspot.com} for the complete list of live test
cases.

\section{Analysis Workflow}

\thesystem automatizes the extraction and analysis of XSS payloads by following
an iterative approach. At a high level, in \acf{phase1} we run one scanner at a
time against our testbed web application, capture the network traffic, and
decode the HTTP requests. Then we extract the payloads from
each request by analyzing the possible sources with a set of heuristics. In
this phase we also retain the vulnerability report produced by each scanner.

We proceed to \acf{phase2}, where in order to simplify the evaluation
and visualization of large volumes of payloads, we apply an algorithm that
clusters the payloads based on their similarity and produces a representative
template that describes the clustered payloads in a compact way. A
\emph{template} is a string composed by lexical tokens (e.g., a parenthesis, a
function name, an angular bracket), common to all the payloads in a cluster, and
variable parts, which we represent with placeholders (e.g., STR, NUM, PUNCT). We
do not claim that our clustering algorithm is accurate, as the problem of
inferring precise templates from strings is unsolved. But we do not need such
precision, as we use this step simply as a mean to reduce the manual review of
the results. Conceptually, the templates are the base payloads employed by a
scanner before applying any mutation or fuzzying.

In \acf{phase3} we evaluate each template on a set of \metrics that quantify
generality, filter evasion capabilities, use of mutations, and type of
triggered vulnerability, presenting them in a compact report with average values
computed on each cluster for ease of visualization.

These steps are executed in an automated fashion by three modules we
implemented in Python (making use of the \texttt{dpkt} library to parse
HTTP requests in Phase 1).

In \acf{phase4}, which cannot be fully automated, we generate a list of
\emph{negative payloads}, i.e., those that are not found in the parsed report.
Using the Template Generator, we discard the redundant negative payloads. For
each resulting distinct negative payload we create specific test cases via
manual reverse-engineering, as outlined in \S\ref{sec:firing-range}.
Iteratively, this results in the reduction of the number of negative payloads to
zero, and the increase of the coverage of each scanner's capabilities.

In the following, we explain in detail the four phases of our approach.

\subsection{\acf{phase1}}
The high-level goal of this phase is to obtain, for each scanner, the entire set
of XSS payloads used by each scanner for each entry point in the testbed
application. To this end, this phase first captures (using \texttt{libpcap}) the
traffic generated by the scanners during the test and performs TCP stream
reassembly, decoding the full HTTP requests.


The first challenge is to automatically distinguish HTTP requests used for the
actual injection (i.e., containing one or more payload) from those used for
crawling or other ancillary functionalities. The ample amount of payloads
generated makes manual approaches unfeasible. Therefore, we rely on two
heuristics:
\begin{description}
\item [Signature Payloads:] Most scanners use ``signature'' payloads (i.e.,
  payloads uniquely used by a scanner). For example, one of the scanners always injects
  payloads that contain a custom, identifying HTML tag ``sfi''. Therefore, we
  derived the signature payloads for each scanner and compiled a whitelist that
  allows this heuristic to discard the uninteresting requests.

\item [Attack Characters:] Since we know the testbed application, we guarantee
  that there is no legitimate request that can possibly contain certain
  characters in the header or body parameters. These characters include
  \verb|<|, \verb|>|, quote, double-quote and their corresponding URL-percent
  encodings, etc. Such characters should not be present in a crawling request by
  construction, and since they are often required to exploit XSS
  vulnerabilities, we have empirically observed them as linked to testing. For
  example, considering the GET request
  \path|.../reflected/body?q=<script>alert(232)</script>|, the value of \path|q|
  contains \path|<|, so this heuristic flags it as a testing request. We apply
  the same approach to POST requests.
\end{description}
To complement the coverage of the previous heuristics and maximize the number of
identified payloads, we perform pairwise comparisons between requests issued by
each couple of scanners. For each couple, we extract the body and URL of the two
requests, and check if they have the same path and the same query parameters. If
so, we compare the values of each query parameter. By construction, Firing Range
provides only a single value for each parameter, thus any mismatch has to be
originated by the scanner fuzzing routine. Once a pair of requests is flagged as
a mismatch, we performed manual inspection to isolate the payload. The number of
such cases is rare enough to make this a manageable process. We iteratively
applied and improved these heuristics until this cross-scanner analysis
generated empty output, and we could confirm through manual inspection that no
more test requests were missed (i.e., all payloads considered).

The astute reader notices that we focus our description on query parameters:
This is part of the design of our testbed, which provides injection points
exclusively on query parameters. Clearly, almost all of the scanners perform
path injection, where the payload is injected in the path section of the
URL. For example, in:
\begin{lstlisting}
  /address<script>prompt(923350)</script>/location/innerHtml
\end{lstlisting}
the payload \path|<script>prompt(923350)</script>| is part of the request
path. This is an important feature for scanners as modern applications often
embed some of their parameters in the path of the request. Although our
extraction process handles this type of entry points, we chose not to analyze
path-based injections. Indeed, manual analysis confirmed that scanners used the
very same set of payloads observed during parameter injection.

\subsection{\acf{phase2}}
\label{sec:payload-templating}
Given the large number payloads generated by each scanner, manually analyzing
and evaluating each of them separately is practically unfeasible. A closer
inspection of the payloads, however, revealed self-evident clusters of similar
payloads. For example, the following payloads:
\begin{lstlisting}
  <ScRiPt >prompt(905188)</ScRiPt>
  <ScRiPt >prompt(900741)</ScRiPt>
\end{lstlisting}
differ only for the value passed to the \texttt{prompt} function. To cluster
similar payloads, inspired by the approach presented in~\cite{Pitsillidis}, we
developed a recursive algorithm for string templating. A \emph{template}, in our
definition, is a string composed by lexical tokens (e.g., a parenthesis, a
function name, an angular bracket), that are common to all the payloads in a
cluster, and variable parts, which we represent with placeholders. The
\texttt{NUM} placeholders replace strings that contains only digits, whereas the
\texttt{STR} placeholders replace strings that contains alphanumeric characters.
For instance, the template for the above example is
\begin{lstlisting}
  <ScRiPt >prompt(90_NUM_)</ScRiPt>
\end{lstlisting}
To generate the templates we leveraged the Levenshtein (or edit) distance (i.e.,
the minimum number of single-character insertions, deletions, or substitutions
required to transform string A to string B).

At each recursion, our algorithm receives as an input a list of strings and
performs a pairwise comparison (without repetition) between elements of the
input list. If the Levenshtein distance between each two compared strings is
lower than a fixed threshold, we extract the matching blocks between the two
strings (i.e., sequences of characters common to both strings). If the length of
all matching blocks is higher than a given threshold, the matches are
accepted. Non-matching blocks are then substituted with the corresponding
placeholders. The output of each recursion is a list of generated templates. All
payloads discarded by the Levenshtein or matching-block thresholding are
appended to the list of output templates, to avoid discarding ``rarer'' payloads
(i.e., outliers) and losing useful samples. The thresholds (maximum Levenshtein
distance and minimum matching block length) are decremented at each cycle by an
oblivion factor, making the algorithm increasingly restrictive. We selected the
parameters of the system, including the oblivion factor, through benchmarks and
empirical experimentation, by minimizing the number of templates missed. This
automatic selection yielded the following values: 20, 0.9 (default case); 20,
0.9 (Acunetix), 15, 0.5 (NetSparker); 15, 0.8 (NTOSpider); 15, 0.9 (Skipfish);
15, 0.5 (W3af). The algorithm stops when a recursion does not yield any new
templates.

For example, considering the following payloads as input
\begin{lstlisting}
  <ScRiPt >prompt(911853)</ScRiPt>
  <ScRiPt >prompt(911967)</ScRiPt>
\end{lstlisting}
the resulting template is
\begin{lstlisting}
  <ScRiPt >prompt(911_NUM_)</ScRiPt>
\end{lstlisting}
whereas for payloads
\begin{lstlisting}
  onerror=prompt("x6haqgl3")>
  onerror=prompt("x6hbcxpn")>
\end{lstlisting}
the resulting template is
\begin{lstlisting}
  onerror=prompt("x6h_STR_")>
\end{lstlisting}

\subsection{\acf{phase3}}
\label{sec:template-evaluation}
We want to assess the quality of payloads in terms of \emph{filter-evasion}
capabilities and amount of \emph{mutations} used by the scanner. Given our
observations above, we apply such evaluation to templates, as opposed to each
single payload.

More precisely, the quality of a template is expressed by the following template
\metrics, which we aggregate as defined in \S\ref{sec:template-evaluation-results}. Note
that the \emph{rationale} behind each \metric is explained on payloads, whereas
the \metric itself is \emph{calculated} on the templates.

\begin{description}
\item [\acf{length}, integer:] The longer a payload is, the easier to spot and
  filter (even by accident). Thus, we calculate the length of each payload
  template to quantify the level of evasion capability.
\item [\acf{charset}, integer:] The presence of particular characters in a
  payload could hit server-side filters, or trigger logging. The presence of a
  character instead of another could reveal an attempt to mutate the string
  (e.g., fuzzing). A symbol can have different meanings depending on the actual
  context. From this rationale we obtain that a payload with a small set of
  characters is ``better'' than one leveraging rare characters.  We calculate
  this \metric on the variable part of each template (i.e., excluding the
  \verb|STR| ad \verb|NUM| tokens).
\item [\acf{callbacks} boolean:] Rather than using standard JavaScript
  functions like \texttt{alert}, a scanner can use custom JavaScript function
  callbacks to bypass simple filters. We interpret this as an evasion attempt.
  If a template contains a function outside the set of built-in JavaScript
  functions, we set this \metric to true.
\item [\acf{encodings}, boolean:] Encoding a payload may let it pass
  unnoticed by some web applications' filters. However, some applications do not
  accept certain encodings, resulting in the application not executing the
  payload. A payload that uses multiple encodings is also more general because,
  in principle, it triggers more state changes in the web application. We set
  this \metric to true if the template contains symbols encoded with a charset
  other than UTF-8 and URL-percent, thus quantifying the level of evasion.
\item [\acf{evasion}, integer:] With this
  \metric we quantify the amount of known techniques to avoid filters in web
  applications. For each template we calculate how many known techniques are
  used by matching against the OWASP
  list\footnote{\url{https://www.owasp.org/index.php/XSS_Filter_Evasion_Cheat_Sheet}}.
\end{description}
Although other \metrics could be designed, we believe that these \metrics are
the bare minimum to characterize a scanner's capabilities and understand more
deeply the quality of the payloads that it produces and process.

%

\subsection{\acf{phase4}}
\label{sec:negative-payloads}
At the end of a scan, each scanner produced a report of the detected
vulnerabilities. We use a report-parsing module that we developed (and released)
for each scanner and correlate the results with the payloads extracted.  In this
way we identify payloads that triggered vulnerabilities, which we call
\emph{positive payloads} and those that did not, called \emph{negative
  payloads}.

We manually verified each negative payload to ensure that it was not
our report-parsing module failing to correlate. We found that there
are at least four reasons for which a negative payload occur:
\begin{itemize}
\item The payload was malformed (e.g., wrong or missing characters, incorrect
  structure) and it was not executed. This is a functional bug in the scanner.
\item The payload was designed for a different context than the one it was
  mistakenly injected in.
\item The scanner used what appears to be the ``right'' payload for the test
  case, but the detection engine somehow failed to detect the exploit.
\item The payload was redundant (i.e., the scanner already discovered a
  vulnerability) in the same location thanks to another payload, and thus will
  not report it again.
\end{itemize}
%
%
Since one of our goals was to create a testbed application as complete as
possible, we wanted to ensure that all negative payloads had a matching test
case in our application. With manual analysis, we proceeded to discard malformed
and redundant payloads from the list of negative payloads. For each remaining
negative payloads we produced a specific vulnerable test case.

To avoid introducing a bias in the results, we crafted each new test case to be
triggered exclusively by the payload type for which it has been created, whereas
the other payloads of the same scanner are rejected, filtered, or escaped. Of
course, nothing would prevent other scanners from detecting the case with a
different payload and that was indeed the intended and expected behavior.

\section{Experimental Results}
\label{sec:experiments}
During our experiments we tested 4 commercial scanners, for which we obtained
dedicated licenses with the support of the vendors (in random order, we used
Acunetix 8.0, NetSparker 3.0.15.0, N-Stalker 10.13.11.28, NTOSpider 6.0.729) and
2 open-source scanners (in random order, we used Skipfish 2.10b and w3af 1.2).

We installed each scanner on a dedicated virtual machine (VM) to guarantee
reproducibility and isolation (i.e., Debian 6.0.7 for Skipfish and W3af, and
Windows 7 Professional for Acunetix, NetSparker and NTOSpider). We used
Wireshark on each VM to capture the traffic. When possible, we configured
each scanner to only look for XSS vulnerabilities, and to minimize the impact of
other variables, we left the configuration to its default values and kept it
unchanged throughout all the tests.

We tested Firing Range several times with each scanner. Overall, the scanners
are generally good at detecting the reflected XSS while performing very poorly
with respect to the DOM based XSS vulnerabilities: none of those included in our
application was identified by any scanner. No scanner reported false positives,
as we were expecting, since we did not design any test cases to trick them like
\citet{Bau2010a} did in their testbed application.

Of course, simply running scanners against a test application and analyzing
their reports is not enough to evaluate their performance. As \citet{Doupe2010}
did in their study, we wanted to understand the behavior of a scanner in action
to be able to explain their results.  Our approach, however, differs noticeably
since \citeauthor{Doupe2010}'s main concern is about the crawling phase of the
scanner, whereas we focus on the attack phase, and specifically on the payloads.

\subsection{\acl{phase1} Results}
\label{sec:payload-extraction-results}
\begin{figure}[t]
  \centering\includegraphics[width=0.7\textwidth]{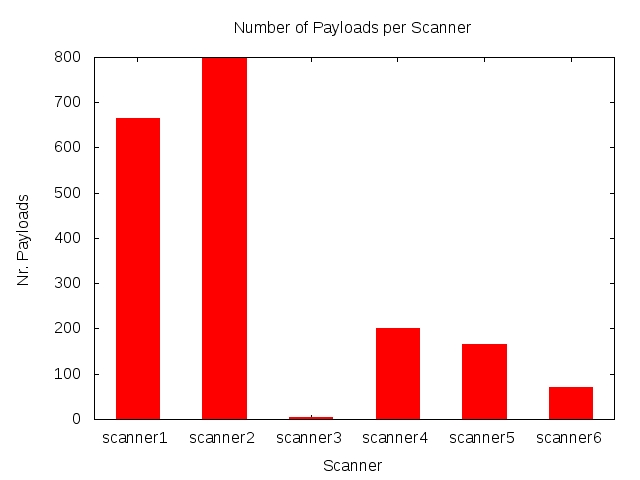}
  \caption{Number of extracted payloads.}
\label{fig:NrPayloads}
\end{figure}

The number of extracted payloads for all scanners is shown in
Fig.~\ref{fig:NrPayloads}.

When looking at the results, it is worth keeping in mind that all scanners
performed very similarly: they were able to detect the same number of
vulnerabilities on the first version of Firing Range, before
\acf{phase4}. Indeed, they all missed the same amount of
vulnerabilities. Although our report-parsing modules are able to obtain these
results automatically, we confirmed this by manual analysis.

With this in mind it is easy to see how a simple measure like the number of
distinct payloads is interesting: the detection technique used by
\textbf{Scanner 3} results in the use of far fewer payloads. The comparatively
larger number of payloads observed in the first 2 scanners is due to the use of
unique identifiers tied to each of requests. We argue that the identifiers are
used to link a result back to the request that originated it even if server side
processing had moved it around---this property is important when detecting
stored XSS.

Fig.~\ref{fig:NrPayloads} also clearly illustrates the benefits of \acf{phase3},
as these numbers do not really tell much about the actual quality and type of
the payloads: it is quite possible that \textbf{Scanner 1} had simply appended
an incremental number to the very same payload.

\subsection{\acl{phase2} Results}
\begin{figure}[t]
  \centering\includegraphics[width=0.7\textwidth]{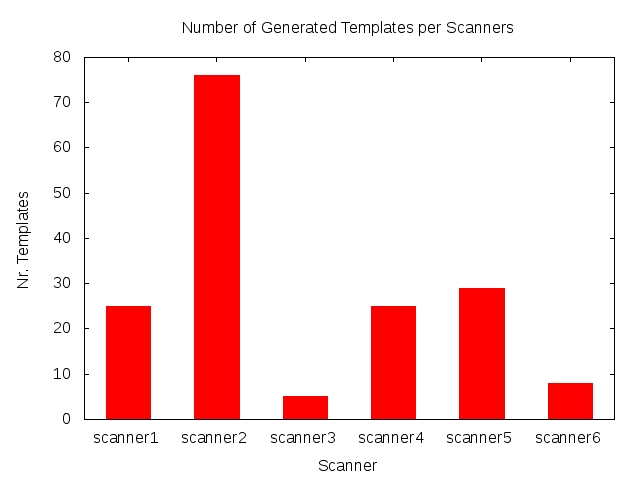}
  \caption{Number of generated templates.}\label{fig:NrTemplates}
\end{figure}

After applying the clustering process described in \acl{phase2}, we notice
immediately the limited number of templates, as shown in
Fig.~\ref{fig:NrTemplates}.

The larger number of templates generated for \textbf{Scanner 2} is an index of
lower efficiency and of lack of generality. These numbers also show that our
hypothesis was correct: the high number of payloads of \textbf{Scanner 1} was
due to an incremental number (or randomly generated number), rather than
structurally different payloads. Indeed, among the templates that we generated
from \textbf{Scanner 1} there is:
\begin{lstlisting}
  -STR-_NUM_<-STR-ScRiPt-STR->prompt(_NUM_)</ScRiPt>-STR-
\end{lstlisting}
where the second \verb|_NUM_| token is the unique identifier.

At this point in the analysis we could already see some scanners emerging as
clearly more efficient due to the smaller number of templates they use. For
example, \textbf{Scanner 2} uses more complex payloads such as:
\begin{lstlisting}
  -STR-`"--></style></script><script>alert(0x0000-STR-_NUM_)</script>
  a\x00`"--></style></script><script>alert(0x0000F8)</script>
\end{lstlisting}
Since the variety of payloads of \textbf{Scanner 3} is very low, also the number
of generated templates is very limited (only 5). The majority, 14, of the
payloads of this scanner are represented by the following template:
\begin{lstlisting}
  -STR-<alert><h1>SCANNER3_XSS-STR-
\end{lstlisting}
where \verb|alert| and \verb|SCANNER3| were replaced to avoid revealing the
scanner's real name. Attempts of evading filters from \textbf{Scanner 3} are
well represented by the following template:
\begin{lstlisting}
  a\r\n\r\n<alert><h1>SCANNER3_XSS
  a><alert><h1>SCANNER3_XSS
\end{lstlisting}
Given the high number and variety of payloads employed by this \textbf{Scanner
  4}, also resulting templates are numerous and very different from each other.
There is one template that represents most of the 200 payloads generated
by this scanner:
\begin{lstlisting}
  -STR-"><script>alert("x6_NUM_-STR-_NUM_")</script>-STR-
\end{lstlisting}
whereas the remainder templates are variations of the above one. For instance:
\begin{lstlisting}
  <img """><script>alert("x6wsuum4")</script>">
  STR<script>alert(/x6STR/)STR</script>
  --></script><script>alert(/x6w2uf13/)</script>
  \r\n\r\n<script>alert(/x6okbc2h/)</script>
\end{lstlisting}
where the latter two template show attempt of evading filters (e.g., multi
line). The (low number of) templates from \textbf{Scanner 5} show that its
payloads are significantly different from the rest. The templates that cover
most of the payloads are:
\begin{lstlisting}
  -->">'>'"<sfi000084v209637>
  .htaccess.aspx-->">'>'"<sfi000085v209637>
  .htaccess.aspx-->">'>'"<sfi000_NUM_v209637>
  -STR--->">'>'"<sfi_NUM_v209637>
\end{lstlisting}
which capture the scanner developer's particular interest in generic
payloads that can highlight the incorrect escaping of a number of special
characters at once. This approach is not found in any other scanner.

\textbf{Scanner 6} created a large number of templates, sign of strong
use of non-trivial mutations and fuzzying. The most interesting samples, which
cluster about 40 payloads each, are:
\begin{lstlisting}
  javas-STR-cript:alert("-STR-");
  javas\x00cript:alert("FSWQ");
  javas\tcript:alert("DmKP");
\end{lstlisting}

\subsection{\acl{phase3} Results}
\label{sec:template-evaluation-results}
During this phase we evaluated each of the templates on the \metrics defined
in \S\ref{sec:template-evaluation}.

\begin{figure}[t]
  \centering\includegraphics[width=0.7\textwidth]{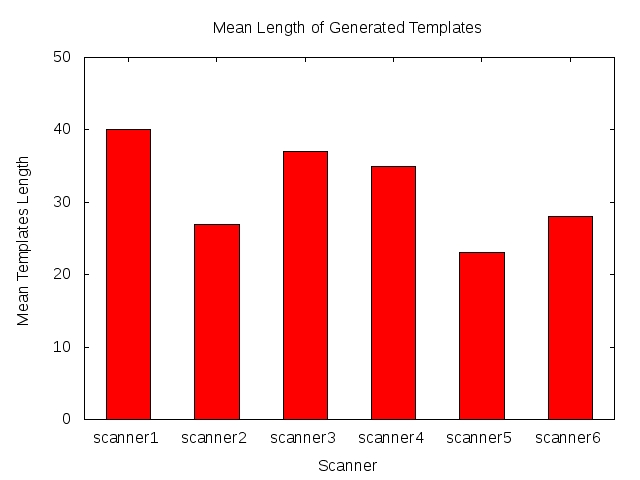}
  \caption{Mean \acf{length} over the templates of each scanner.}
  \label{fig:TemplLength}
\end{figure}

Fig.~\ref{fig:TemplLength} reports the mean of \acf{length} calculated over the
number of templates produced by each scanner. This is an interesting finding,
which can be interpreted in two ways. On the one side, the length of the
templates is in line with the minimum length of real-world payloads required to
exploit XSS vulnerabilities, which is around 30
characters~\cite{gnarlysec,shortening}, which somehow justifies the choice of
the payloads. On the other hand, 30 characters is also quite a long
string. Indeed, if a vulnerable entry point (e.g., \path|id=311337|) that
employs a weak sanitization filter (e.g., based on a cut-off length) will not be
detected as vulnerable by these long payloads. Although this can be a good way
for the scanner to avoid flagging unexploitable vulnerabilities (false
positives), it has been shown that multiple small payloads can be combined to
generate a full attack~\cite{highseverity}. However, the scanners that we
examined miss these occurrences.

\begin{table}[b]
  \caption{Summary of template evaluation.}
  \label{table:TemplSumm}
  \centering
  \begin{tabular}{lccc}
    \toprule
    \textsc{Scanner} & \textsc{Mutations (\ac{encodings})} & \textsc{Callbacks (\ac{callbacks})} & \textsc{Filter evasion (\ac{charset}, \ac{encodings}, \ac{evasion})}\\
    \midrule
    \textbf{Scanner 1} & \OK & \KO & \OK\\
    \textbf{Scanner 2} & \OK & \OK & \OK\\
    \textbf{Scanner 3} & \OK & \OK & \OK\\
    \textbf{Scanner 4} & \OK & \KO & \OK\\
    \textbf{Scanner 5} & \KO & \KO & \KO\\
    \textbf{Scanner 6} & \OK & \KO & \OK\\
    \bottomrule
  \end{tabular}
\end{table}

We notice that \textbf{Scanner 1} employs significantly longer payloads than
\textbf{Scanner 5}. This can be explained, considering that \textbf{Scanner 5}'s
\ac{evasion} is zero, meaning that it uses no known filter-evasion techniques:
thus, \textbf{Scanner 5} is less sophisticated than \textbf{Scanner 1}.

Using \ac{charset}--\ac{evasion}, we derived Table~\ref{table:TemplSumm}, which
gives bird's eye view on the use of mutations, filter evasion and use of
callbacks from each scanner. Regarding callbacks and mutations, we use
\acf{callbacks} and \acf{encodings}, whereas for filter evasion, if at least one
template has a non empty character set (from \ac{charset}), uses multiple
encodings (from \ac{encodings}), and adopt at least one evasion technique (from
\ac{evasion}) we consider the scanner as using filter evasion.

As it can be seen, both random mutations and filter-evasion techniques are
widely employed in the scanners that we tested. We comment on their
effectiveness in \S\ref{sec:negative-payloads-results}, but our analysis
suggests that such features are now widespread. The use of callbacks over
parsing or standard JavaScript functions, on the other hand, is not very used.

\subsection{\acl{phase4} Results}
\label{sec:negative-payloads-results}
As described in \S\ref{sec:firing-range} we have iteratively added new test
cases in our testbed to account for negative payloads. The expected result was
the reduction of the number of negative payloads to zero.

\begin{figure}[t]
  \centering\includegraphics[width=0.7\textwidth]{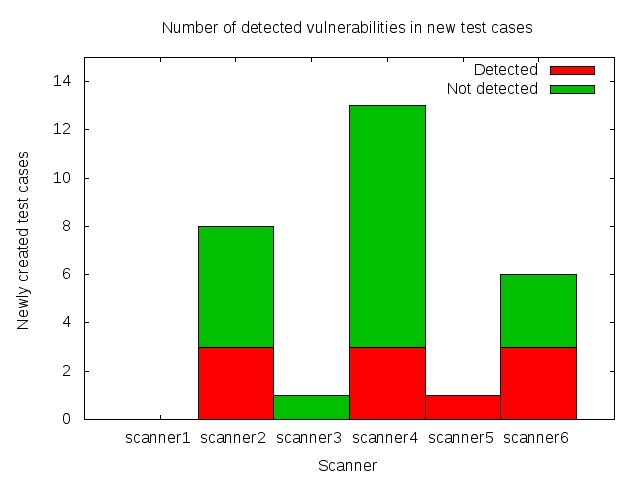}
  \caption{Results on the new test cases alone.}
  \label{fig:NegativePayloadResults}
\end{figure}

We ran all the scanners against the new testbed and analyzed the
results. Unfortunately, as Fig.~\ref{fig:NegativePayloadResults} shows, scanners
failed to detect most of the new vulnerabilities. We manually manually confirmed
that the negative payloads do trigger an XSS on those new cases. However, the
very same scanners that generated such negative payloads still failed to detect
and report most of the new vulnerabilities. Note that the payloads of
\textbf{Scanner 1} were all already covered by our initial test case, thus no
additional cases were created.

\begin{figure}[t]
  \centering\includegraphics[width=0.7\textwidth]{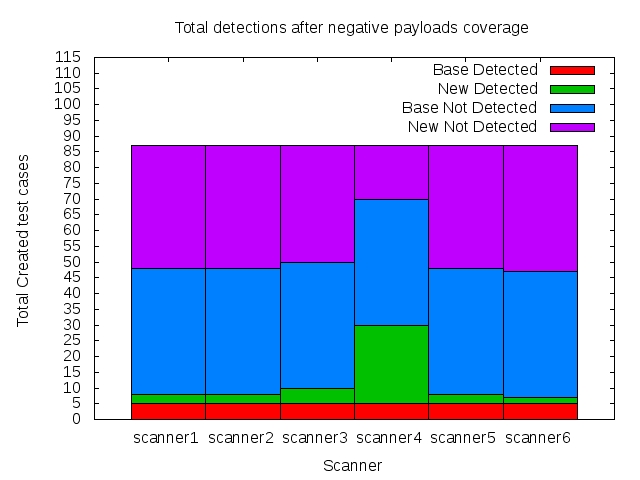}
  \caption{Final results after \ac{phase4}, including ``New'' test
    cases. ``Base'' are the initial test cases, ``Detected'' are true positive,
    whereas ``Not Detected'' are false negative.}
  \label{fig:FinalResults}
\end{figure}

Fig.~\ref{fig:FinalResults} shows the overall results produced by each scanner
after including the new test cases.

By manually analyzing the requests toward the new cases, we discovered that some
scanners did not exercise the test case with the right payload although they
should have been able to: A faulty (or random) payload-selection procedure
somehow failed to choose the right payload, instead using it in test cases where
it would turn out to be ineffective.

Another interesting result is that, after the introduction of the new cases,
some scanners started using payloads we had not observed before. This behavior
suggests some degree of context awareness, as scanners would only generate this
new set of templates after having observed these new contexts. However, even in
this case we observed a staggering high rate of failures for the new corner
cases we added.

Although scanners did not achieve the expected results, this process allowed us
to greatly increase our coverage of test cases for attacks supported by the
analyzed scanners, and to produce a state of the art testbed for future work.

\section{Related Work}
\label{sec:related-work}
As explained in \S\ref{sec:introduction} and \ref{sec:background}, most of the
state of the art and related work differ from ours in the approach, analysis
perspective (payload), and scope (XXS in depth). Although we also evaluate the
scanner detection rate, we focus on extracting the payloads for analyzing
them. While doing this, we try not to trick the scanner into complex contexts,
because our goal is to extract as many payloads as possible.

The most closely related work is by~\citet{Doupe2010}, who tested 11 commercial
and open-source scanners on a realistic testbed application called
``WackoPicko'', containing some of the most popular and known vulnerabilities
(e.g., XSS, SQL injections, parameter manipulation) and challenges for scanner
crawler modules. In their application they included reflected and stored XSS,
without including DOM, which was one of our main concerns. They also performed
comparisons between scanners by arranging them in a lattice ordered based on
detected vulnerabilities, differently from our work, which is not
comparative. In their evaluation, they try to explain scanners' results by
focusing on the crawling phase (rather than the attack phase as we did) and
underline the importance of crawling challenges.

Another closely related work is by~\citet{Bau2010a}, which evaluates 8 black-box
web vulnerability scanners, on the parameters of the supported classes of
vulnerabilities, effectiveness against target vulnerabilities and how
representative generated tests were, when compared to real world
vulnerabilities. They also implemented a custom web application, containing
known vulnerabilities, to use as a target. Their application also includes traps
designed to trick scanners into reporting false positives. Their perspective
and focus, however, is completely different by ours.

Another work by~\citet{Doup} takes into consideration limitations in interacting
with complex applications, due to the presence of multiple actions that can
change the state of an application. They propose a method to infer the
application internal state machine by navigating through it, observing
differences in output and incrementally producing a model representing its
state. They then employ the internal state machine to drive the scanner in
finding and fuzzing input vectors to discover vulnerabilities. To evaluate the
approach, they ran their state-aware scanner along with three other
vulnerability scanners, using as metrics real total detections, false positives
and code coverage. Like the previously cited work, \cite{Doup} differs
from our approach in the focus of the analysis.

Several other works perform evaluation of web vulnerability
scanners. \citet{Vieira2009} tested four web scanners on 300 web services,
\citet{khoury2011} evaluated three state of art black-box scanners supporting
detection of stored SQL injection vulnerabilities on a custom testbed
application, \citet{chen2010} conducted an evaluation of 40 popular scanners on
several test applications containing XSS vulnerabilities. However, as previously
mentioned, previous work was not XSS specific, and not focused on payloads.

\citet{Suto2010a,suto2007} evaluate the accuracy and time needed to run and
review and supplement the results of 7 web application scanners. To test
scanners they did not use custom testbed applications, opting for
vendor-provided ones, which could however lead to biased results. This type of
work complement ours and the previously cited ones, because they focus on
pure performance indicators rather than (only) on detection capabilities.

\section{Conclusions}
\label{sec:concl}

Our analysis, the first vertical study on XSS vulnerability scanners, produced
quality \metrics of 6 commercial and open-source products through passive
reverse engineering of their testing phases, and manual and automated analysis
of their payloads. Furthermore, we created a reliable, reusable and publicly
available testbed.

Our results reveal a high variance in the number of distinct payloads used by
each scanner, with numbers ranging from an handful to over 800, as well as in
the variety of payload types: the scanners generate from a couple to tenths of
unique payload types. The first element is significant when analyzed in
conjunction with the number of detected vulnerabilities, because it shows the
redundancy and lack of efficiency of some scanners.

We also analyzed the structure of payload templates to assess their quality
using a set of specific \metrics. This analysis highlighted remarkable
discrepancies in the type and structure of payloads: some scanners leverage
complex and advanced attack strings designed to trigger vulnerabilities in rare
corner cases, others use basic, versatile payloads.

Notably, none of the scanners we assessed was capable of detecting DOM XSS.

Finally, by iterating on payloads that triggered no test cases, we were able to
noticeably improve our test application and draw important conclusions about
each scanner's inside workings. One of the key results is that, despite having
some kind of awareness about context, all of the tested scanners were found
wanting in terms of selecting the attack payloads and optimizing the number of
requests produced. A staggering high number of detection failures suggest bugs
and instability in the detection engines, while the high variance in types
and features of the payloads we inspected makes the case for cooperation in
defining common, efficient and reliable payloads and detection
techniques.\todo{Sto provando a chiudere con una call for action che ha sempre presa. Opinioni?}

\bibliographystyle{splncsnat.bst} 
\bibliography{xss-peeker-paper}

\begin{thebibliography}{26}
\providecommand{\natexlab}[1]{#1}
\providecommand{\url}[1]{\texttt{#1}}
\providecommand{\urlprefix}{}

\bibitem[{Bau et~al.(2010)Bau, Bursztein, Gupta, and Mitchell}]{Bau2010a}
Bau, J., Bursztein, E., Gupta, D., Mitchell, J.: State of the art: Automated
  black-box web application vulnerability testing.
\newblock In: {IEEE} SSP. pp. 332--345 (May 2010)

\bibitem[{Cenzic(2013)}]{Cenzic}
Cenzic: {Cenzic Application Vulnerability Trends Report} (2013),
  \urlprefix\url{http://info.cenzic.com/rs/cenzic/images/Cenzic-Application-Vulnerability-Trends-Report-2013.pdf}

\bibitem[{Chen(2010)}]{chen2010}
Chen, S.: Web application scanners accuracy assessment (2010),
  \urlprefix\url{http://sectooladdict.blogspot.com/2010/12/web-application-scanner-benchmark.html}

\bibitem[{Doup{\'e} et~al.(2012)Doup{\'e}, Cavedon, Kruegel, and Vigna}]{Doup}
Doup{\'e}, A., Cavedon, L., Kruegel, C., Vigna, G.: Enemy of the state: A
  state-aware black-box web vulnerability scanner.
\newblock In: {USENIX} Security. pp. 26--26. USENIX Association (2012)

\bibitem[{Doup{\'e} et~al.(2010)Doup{\'e}, Cova, and Vigna}]{Doupe2010}
Doup{\'e}, A., Cova, M., Vigna, G.: Why johnny can't pentest: An analysis of
  black-box web vulnerability scanners.
\newblock In: DIMVA. pp. 111--131. Springer-Verlag, Berlin, Heidelberg (2010)

\bibitem[{Foundstone(2006)}]{hacmeBank}
Foundstone: {Hacme Bank v2.0} (2006),
  \urlprefix\url{http://www.foundstone.com/us/resources/proddesc/hacmebank.html}

\bibitem[{Gnarlysec(2010)}]{gnarlysec}
Gnarlysec: {XSS and ultra short URLs} (2010),
  \urlprefix\url{http://gnarlysec.blogspot.ch/2010/01/xss-and-ultra-short-urls.html}

\bibitem[{Heiderich et~al.(2013)Heiderich, Schwenk, Frosch, Magazinius, and
  Yang}]{Heiderich2013}
Heiderich, M., Schwenk, J., Frosch, T., Magazinius, J., Yang, E.Z.: {mXSS}
  attacks: Attacking well-secured web-applications by using {innerHTML}
  mutations.
\newblock In: {CCS}. pp. 777--788. ACM (2013)

\bibitem[{Khoury et~al.(2011)Khoury, Zavarsky, Lindskog, and Ruhl}]{khoury2011}
Khoury, N., Zavarsky, P., Lindskog, D., Ruhl, R.: An analysis of black-box web
  application security scanners against stored sql injection.
\newblock In: {IEEE} 3rd International Conference on Privacy, security, risk
  and trust (PASSAT). pp. 1095--1101 (Oct 2011)

\bibitem[{Klein(2005)}]{klein2005dom}
Klein, A.: {DOM} based cross site scripting or {XSS} of the third kind (2005),
  \urlprefix\url{http://www.webappsec.org/projects/articles/071105.shtml}

\bibitem[{Lekies et~al.(2013)Lekies, Stock, and Johns}]{Lekies2002}
Lekies, S., Stock, B., Johns, M.: 25 million flows later: Large-scale detection
  of dom-based xss.
\newblock In: CCS. pp. 1193--1204. ACM (2013)

\bibitem[{Maggi et~al.(2013)Maggi, Frossi, Zanero, Stringhini, Stone-Gross,
  Kruegel, and Vigna}]{longshore}
Maggi, F., Frossi, A., Zanero, S., Stringhini, G., Stone-Gross, B., Kruegel,
  C., Vigna, G.: Two years of short urls internet measurement: Security threats
  and countermeasures.
\newblock In: WWW. pp. 861--872 (2013)

\bibitem[{Mutton(2011)}]{highseverity}
Mutton, P.: {XSS in confined spaces} (2011),
  \urlprefix\url{http://www.highseverity.com/2011/06/xss-in-confined-spaces.html}

\bibitem[{{Open Web Application Security Project}(2008)}]{owaspTestingGuide}
{Open Web Application Security Project}: {Testing Guide v3} (2008),
  \urlprefix\url{https://www.owasp.org/images/5/56/OWASP_Testing_Guide_v3.pdf}

\bibitem[{{Open Web Application Security
  Project}(2013{\natexlab{a}})}]{xssTop10}
{Open Web Application Security Project}: Top ten (2013{\natexlab{a}}),
  \urlprefix\url{https://www.owasp.org/index.php/Top_10_2013-Top_10}

\bibitem[{{Open Web Application Security
  Project}(2013{\natexlab{b}})}]{webGoat}
{Open Web Application Security Project}: {WebGoat Project}
  (2013{\natexlab{b}}),
  \urlprefix\url{http://www.owasp.org/index.php/Category:OWASP WebGoat Project}

\bibitem[{Pitsillidis et~al.(2010)Pitsillidis, Levchenko, Kreibich, Kanich,
  Voelker, Paxson, Weaver, and Savage}]{Pitsillidis}
Pitsillidis, A., Levchenko, K., Kreibich, C., Kanich, C., Voelker, G.M.,
  Paxson, V., Weaver, N., Savage, S.: Botnet judo: Fighting spam with itself.
\newblock In: DNSS. The Internet Society, San Diego, California, USA (March
  2010)

\bibitem[{Research(2013)}]{msrcXss}
Research, M.: Microsoft security intelligence report (cross-site scripting)
  (2013),
  \urlprefix\url{http://www.microsoft.com/security/sir/strategy/default.aspx#!xss_trends}

\bibitem[{Samuel et~al.(2011)Samuel, Saxena, and Song}]{Samuel}
Samuel, M., Saxena, P., Song, D.: Context-sensitive auto-sanitization in web
  templating languages using type qualifiers.
\newblock In: CCS. pp. 587--600. ACM (2011)

\bibitem[{SecureWorks(2012)}]{dell}
SecureWorks, D.: {Dell SecureWorks Threat Report for 2012} (2012),
  \urlprefix\url{http://www.secureworks.com/cyber-threat-intelligence/threats/2012-threat-reviews}

\bibitem[{Suto(2007)}]{suto2007}
Suto, L.: {Analyzing the Effectiveness and Coverage of Web Application Security
  Scanners} (2007),
  \urlprefix\url{http://www.ntobjectives.com/files/CoverageOfWebAppScanners.pdf}

\bibitem[{Suto(2010)}]{Suto2010a}
Suto, L.: {Analyzing the accuracy and time costs of web application security
  scanners} (2010),
  \urlprefix\url{http://www.ntobjectives.com/files/Accuracy_and_Time_Costs_of_Web_App_Scanners.pdf}

\bibitem[{Toews(2012)}]{shortening}
Toews, B.: {XSS shortening cheatsheet} (2012),
  \urlprefix\url{http://labs.neohapsis.com/2012/04/19/xss-shortening-cheatsheet}

\bibitem[{Tudor(2013)}]{Tudor2013}
Tudor, J.: {Web Application Vulnerability Statistics 2013}.
\newblock available online at \url{
  http://www.contextis.com/files/Web_Application_Vulnerability_Statistics_-_June_2
  013.pdf} (June 2013)

\bibitem[{Vieira et~al.(2009)Vieira, Antunes, and Madeira}]{Vieira2009}
Vieira, M., Antunes, N., Madeira, H.: Using web security scanners to detect
  vulnerabilities in web services.
\newblock In: IEEE/IFIP DSN. pp. 566--571 (June 2009)

\bibitem[{X-Force(2013)}]{xforce}
X-Force, I.: {IBM X-Force 2013 Mid-Year Trend and Risk Report} (2013),
  \urlprefix\url{http://securityintelligence.com/cyber-attacks-research-reveals-top-tactics-xforce}

\end{thebibliography}

\end{document}